\documentclass[aps,english,showpacs,twocolumn,pra]{revtex4-1}
\usepackage{amsfonts}
\usepackage{amssymb}
\usepackage{amsmath}
\usepackage{graphicx}
\usepackage{epsfig}

\begin{document}

\title{Flat band in two-dimensional non-Hermitian optical lattices}
\author{S. M. Zhang}
\author{L. Jin}
\email{jinliang@nankai.edu.cn}
\affiliation{School of Physics, Nankai University, Tianjin 300071, China}

\begin{abstract}
We propose a method to generate a real-energy flat band in a two-dimensional
(2D) non-Hermitian Lieb lattice. The coincidence of the flat band eigenstate
in both real and momentum spaces is essential for the proposed flat band,
which is flexible at the appropriate match between the synthetic magnetic
flux and non-Hermiticity. The proposed method is not limited to the 2D
non-Hermitian Lieb lattice, and is applied to the 2D non-Hermitian Tasaki's
decorated square lattice, dice lattice, and kagome lattice. Our findings
make a step forward for the flat band engineering in 2D non-Hermitian
optical lattices.
\end{abstract}

\maketitle

\section{Introduction}

The flat band is a dispersiveless energy band with fixed energy \cite{Vidal}%
. It is independent of the system momentum and leads to unconventional
Anderson localization \cite{Anderson,Shukla10} and Aharonov-Bohm cages \cite%
{SLOL,JC,JVidal,PRL18,ASarXiv}. The superposition of flat band eigenstates
exhibits no diffraction dynamical behavior. Systems with flat bands have
been a hot topic in the optical study \cite%
{LiebNJP,Scaling,LGeLL,VA,MolinaPRL,Julku,SFlachPRL113,ThomsonPRL2015,Bermudez,Yamamoto,Ventra,Wu,Shukla18,LeykamFB,LGePR,LGePRL2018}%
. Synthetic magnetic flux in optical systems \cite%
{Yu,KFang,ELi,Roushan,Hafezi,HafeziPRL14} enables the destructive
interference, resulting in a flat band. The eigenstates of the flat band are
compact localized states (CLSs) \cite%
{FlachPRB17,LeykamAPX,Hsu,CTChanBIC,SFlachPRB95,SFlachEPL,Nori}, and the
excitations of CLSs are confined to one or more unit cells.

Flat bands have been proposed and experimentally realized in a large number
of two-dimensional (2D) lattices, including the 2D Lieb lattice \cite%
{LiebNJP,MolinaPRL,ThomsonPRL2015}, kagome lattice \cite{Yamamoto,Ventra},
honeycomb lattice \cite{Wu}, dice lattice \cite{Kolovsky}, and others \cite%
{Xiaopeng Li,LeykamAPX}. The Lieb lattice has a three-site unit cell; the
chiral symmetry ensures energy spectra corresponding to the flat band with a
zero energy and two symmetric dispersive bands, with the three energy
spectra sharing a common Dirac point \cite{Lieb}. The chiral flat band in a
generalized three-dimensional (3D) Lieb lattice has also been verified to
possess the CLSs \cite{FlachPRB17}. In addition, other one-dimensional (1D)
models such as the Lieb lattice \cite{Baboux}, rhombic lattice \cite%
{Bermudez,OLMukherjee,FlachBO}, cross-stitch lattice \cite%
{SFlachPRL113,Nori,Huber}, and triangular lattice \cite{Shukla18} support
flat bands.

Parity-time ($\mathcal{PT}$) symmetry in non-Hermitian systems \cite%
{Bender98,Ruschhaupt,NM,El}, in particular, a coupled $\mathcal{PT}$%
-symmetric dimer system \cite{AGuo,CE,BP,LChang,Hodaei,WChen}, has been
widely investigated in various physical aspects \cite%
{MalomedOL,CPA,LFeng,Hodaei17,LFengScience,Assawaworrarit,Schomerus,LJin,Joglekar,CHLee,Zhu,Alu,ZZhang,SHFan,PTRevLGe,Kominis,PTRev,LJinPRL}%
. including coherent perfect absorption \cite{CPA}, unidirectional
invisibility \cite{LFeng}, wireless energy transfer \cite{Assawaworrarit},
unidirectional propagation and unidirectional lasing \cite{LJinPRL}.
Topological photonic crystals with global non-Hermiticity have also been
realized in experiments \cite%
{Zeuner,KDing,Segev2011,PTRevAlu,Longhi18,Xu,YangPNAS,Harari}. These provide
an optical platform for the study of flat bands in non-Hermitian systems. In
practice, flat bands appear in a large number of non-Hermitian systems \cite%
{LeykamFB,JLFB,LGePRL2018,LGeLieb,Molina,RamezaniFB}. Non-Hermitian systems
with the combination of a bipartite symmetry and frustration support flat
bands \cite{LeykamFB}. The flat band maintains at the appropriate match
between the synthetic magnetic flux and non-Hermiticity under destructive
interference \cite{JLFB}. With the precondition of non-Hermitian
particle-hole symmetry, photonic zero modes in a flat band can appear in the
gain and loss modulated lattices \cite{LGePRL2018}. In the quasi 1D cases,
several types of lattices possess flat bands with the additional $\mathcal{PT%
}$-symmetric gain and loss \cite{Molina}; examples include the 1D Lieb
lattice \cite{LGeLieb}. A flat band and light localization can be
manipulated in a non-Hermitian $\mathcal{PT}$-symmetric lattice \cite%
{RamezaniFB}.

In this study, we propose generation of flat bands in 2D non-Hermitian
lattices. In the non-Hermitian generalization, a flat band in the Lieb
lattice is maintained due to the destructive interference. The precondition
for a tunable flat band is that the flat-band eigenstate in the momentum
space is in accordance with that in the real space. The tunable flat-band
energy, whose values are real numbers, is manipulated in terms of variation
of system parameters, including the synthetic magnetic flux caused by a
nonreciprocal phase factor and non-Hermiticity. The position where a flat
band appears in energy spectra is changeable and it intersects with one of
the other two dispersive bands, except in the chiral symmetry condition.
Notably, exceptional points (EPs) are double degenerate under the
non-Hermitian condition. CLSs exist in such a model, and the time evolution
of the initial excitation is stable as time increases. Furthermore, a
simple\ popularization of the construction method of adjustable flat bands
is also provided for other 2D lattice, including the Tasaki's decorated
square lattice, dice lattice and kagome lattice.

The remainder of this paper is organized as follows. In Sec. \ref{II}, we
introduce a 2D non-Hermitian Lieb lattice. The method to achieve a tunable
flat band and the energy band structure are introduced in Sec. \ref{III} and
Sec. \ref{IV}, respectively. CLSs are depicted in Sec. \ref{V}. In Sec. \ref%
{VI}, generalization of the procedure to achieve a tunable flat band for
other 2D lattices is introduced. The summary is presented in Sec. \ref{VII}.

\section{Non-Hermitian 2D Lieb Lattice}

\label{II}

\begin{figure}[t]
\includegraphics[bb=12 242 430 730,  width=6.7 cm, clip]{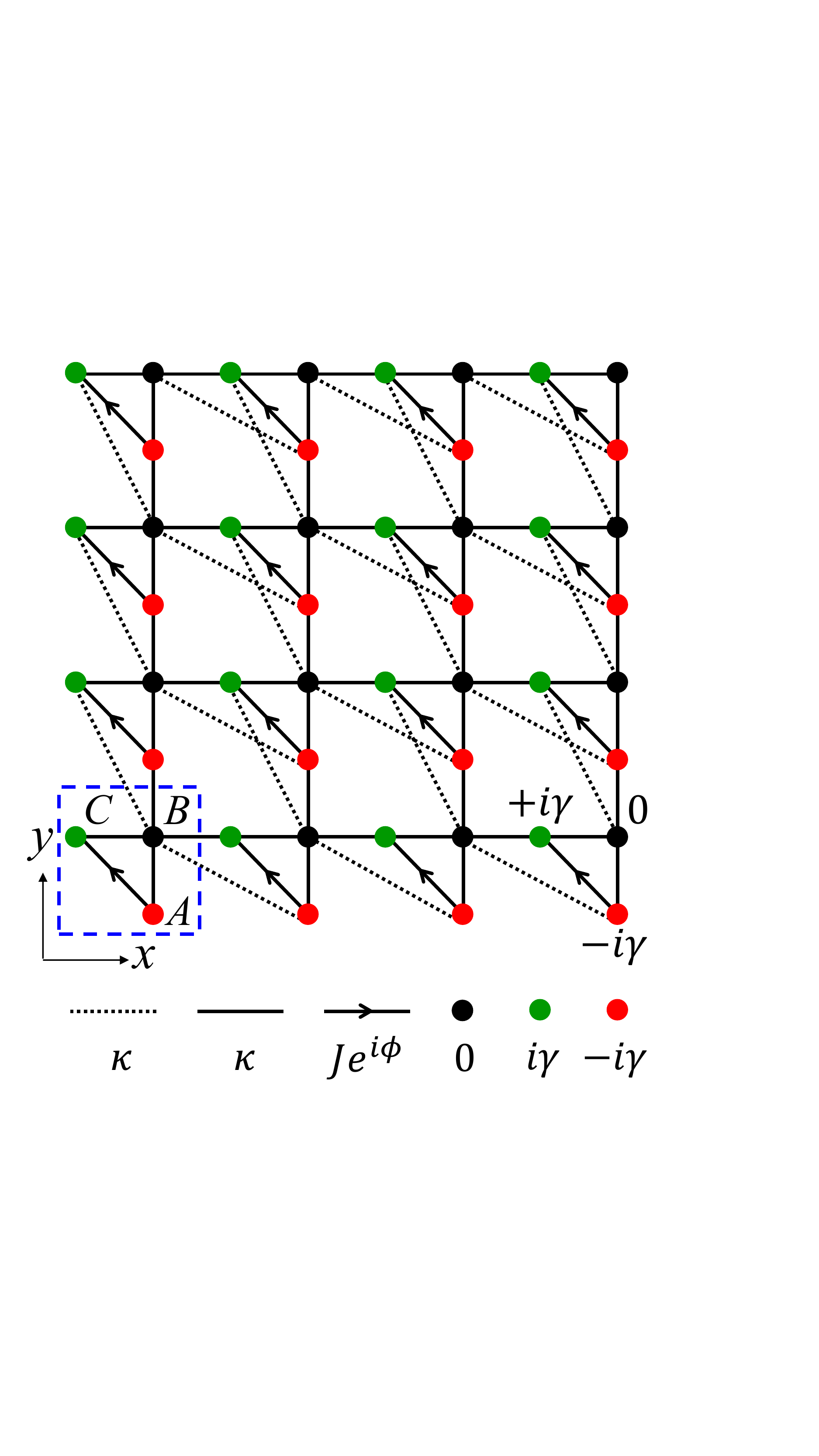}
\caption{Schematic of the two-dimensional Lieb lattice. The sites with gain
(loss) are in green (red), and the sites in black are passive. The
black arrow represents the nonreciprocal coupling, which is $J e^{i\protect\phi}$ ($J e^{-i\protect\phi%
} $) for photons tunneling along (against) the arrow. All the dotted and
solid black lines indicate the reciprocal coupling $\protect%
\kappa $. A unit cell involving sublattices $A$, $B$, and $C$ is exhibited
by the blue dashed box.}
\label{fig1}
\end{figure}

In the non-Hermitian generalization, $A$, $B$, and $C$ represent different
types of sublattices. $A_{j}$, $B_{j}$, and $C_{j}$ constitute the $j$th
unit cell, depicted by the blue dashed box in Fig. \ref{fig1}. The
non-Hermitian Hamiltonian in the real space is given by%
\begin{eqnarray}
H &=&\sum_{m,n}[\kappa b_{m,n}^{\dagger }(a_{m,n}+a_{m,n+1}+a_{m+1,n})
\notag \\
&&+\kappa b_{m,n}^{\dagger }(c_{m,n}+c_{m+1,n}+c_{m,n+1})]  \notag \\
&&+Je^{i\phi }c_{m,n}^{\dagger }a_{m,n}+\mathrm{H.c.}  \notag \\
&&+i\gamma (-a_{m,n}^{\dagger }a_{m,n}+c_{m,n}^{\dagger }c_{m,n}),
\end{eqnarray}%
where $a_{m,n}^{\dagger }$ and $a_{m,n}$ as well as $b_{m,n}^{\dagger }$ $%
\left( c_{m,n}^{\dagger }\right) $ and $b_{m,n}$ $\left( c_{m,n}\right) $
are the creation and annihilation operators, respectively. The parameter $%
\kappa $ represents the coupling strength between different sites, which can
be deemed as positive real numbers without loss of generality. The
nonreciprocal coupling $Je^{\pm i\phi }$ between lattices $A$ and $C$
means that the coupling amplitude is $J$, but its phase factor depends on the direction of photons tunnelling. Photons tunneling from $A$ to $%
C $ will acquire a Peierls phase factor $e^{i\phi }$ \cite%
{Hafezi,HafeziPRL14}; in contrast, photons tunneling oppositely from $C$ to $A$ will
acquire a Peierls phase factor $e^{-i\phi }$ \cite{JL1618}. The gain and
loss rates are $\gamma $.

Under the periodic boundary condition, by taking the Fourier transformation $%
\rho _{m,n}^{\dagger }=(MN)^{-1/2}\sum_{\mathbf{k}}e^{-i(k_{x}m+k_{y}n)}\rho
_{\mathbf{k}}^{\dagger }$ ($\rho =a,b,c$), we obtain $H=\sum_{\mathbf{k}}H_{%
\mathbf{k}}$, where the wave vector $k_{x}=2m\pi /M$, $k_{y}=2n\pi /N$, with
$m$ and $n$ both being integers that range from 1 to $M$ and $N$,
respectively. $H_{\mathbf{k}}$ is the Hamiltonian in the momentum space.
Under the basis $\{a_{\mathbf{k}}^{\dagger }\left\vert \mathrm{vac}%
\right\rangle $, $b_{\mathbf{k}}^{\dagger }\left\vert \mathrm{vac}%
\right\rangle $, $c_{\mathbf{k}}^{\dagger }\left\vert \mathrm{vac}%
\right\rangle \}$, $H_{\mathbf{k}}$ is a $3\times 3$ matrix in the form
\begin{equation}
H_{\mathbf{k}}=\left(
\begin{array}{ccc}
-i\gamma & \kappa \Lambda ^{\dagger }\left( \mathbf{k}\right) & Je^{-i\phi }
\\
\kappa \Lambda \left( \mathbf{k}\right) & 0 & \kappa \Lambda \left( \mathbf{k%
}\right) \\
Je^{i\phi } & \kappa \Lambda ^{\dagger }\left( \mathbf{k}\right) & i\gamma%
\end{array}%
\right) ,
\end{equation}%
where $\Lambda \left( \mathbf{k}\right) =1+e^{ik_{x}}+e^{ik_{y}}$. The 2D
Lieb lattice maintains time-reversal symmetry, $\mathcal{T}H_{\mathbf{k}}%
\mathcal{T}^{-1}=H_{-\mathbf{k}}$, where the time-reversal operator $%
\mathcal{T}=R\mathcal{K}$ with the unitary operator $R$ and the complex
conjugation operator $\mathcal{K}$. When $J=0$ or $\phi =n\pi +\pi /2$ ($n\in\mathbb{Z}$),
the Hamiltonian in the momentum space satisfies the chiral symmetry, $CH_{\mathbf{k}}C^{-1}=-H_{\mathbf{k}}$. The unitary operator $R$ and
the chiral operator $C$ are defined as
\begin{equation}
R=\left(
\begin{array}{ccc}
0 & 0 & 1 \\
0 & 1 & 0 \\
1 & 0 & 0%
\end{array}%
\right) ,C=\left(
\begin{array}{ccc}
0 & 0 & 1 \\
0 & -1 & 0 \\
1 & 0 & 0%
\end{array}%
\right) .
\end{equation}

From the determinant of the matrix $H_{\mathbf{k}}-E_{\mathbf{k}}I_{3\times
3}$ ($I_{3\times 3}$ is the $3\times 3$ identity matrix), that is \textrm{det%
}$\left( H_{\mathbf{k}}-E_{\mathbf{k}}I_{3\times 3}\right) =0$, we obtain a
cubic equation of $E_{\mathbf{k}}$
\begin{equation}
E_{\mathbf{k}}^{3}-\left( 2\kappa ^{2}s_{\mathbf{k}}+J^{2}-\gamma
^{2}\right) E_{\mathbf{k}}-2\kappa ^{2}Js_{\mathbf{k}}\cos \phi =0,
\label{Cubic}
\end{equation}%
where $s_{\mathbf{k}}=2\cos \left( k_{x}-k_{y}\right) +2\left( \cos
k_{x}+\cos k_{y}\right) +3$. Notably, the parameter $s_{\mathbf{k}}$ is
within the range of $s_{\mathbf{k}}\in \lbrack 0,9]$. The energy bands can
be analytically obtained from the solution of the cubic equation.

In the following section, we discuss the energy bands of the non-Hermitian
Lieb lattice. We show that (i) a zero-energy flat band presents in the
lattice with chiral symmetry; (ii) the flat band
persists at an appropriate match between the nonreciprocal coupling and
non-Hermiticity in the absence of chiral symmetry. We are not interested in the trivial situation ($\kappa =0$), where all unit cells are isolated.

\section{Flat Band}

\label{III}

Under chiral symmetry, the system spectrum is symmetric about zero
energy. A three-band chiral symmetric system must have a zero-energy flat band and two dispersive bands with opposite energies.
Equation (\ref{Cubic}) reduces into $E_{\mathbf{k}}\left( E_{%
\mathbf{k}}^{2}+\gamma ^{2}-2\kappa ^{2}s_{\mathbf{k}}\right) =0$ for $J=0$
or $E_{\mathbf{k} }\left( E_{\mathbf{k}}^{2}+\gamma ^{2}-J^{2}-2\kappa
^{2}s_{\mathbf{k} }\right) $$=0$ for $\phi =n\pi +\pi /2$ ($n\in
\mathbb{Z}
$). Thus, a zero-energy flat band with  $E_{\mathbf{k}}=0$ is protected by the
chiral symmetry.

The original 2D Lieb lattice without non-Hermiticity and synthetic magnetic
flux supports a zero-energy flat band \cite{LiebNJP}, that is, only the
nearest neighbour couplings [solid black lines in Fig.~\ref{fig1}(a)] present in $%
H_{\mathbf{k}}$. The three energy bands are given by $0$, $\pm \kappa \sqrt{%
2\cos k_{x}+2\cos k_{y}+4}$. The zero-energy band is independent of the
momentum, and thus, it is a dispersionless flat band. The eigenstate in the
momentum space for the zero-energy flat band as a result of destructive
interference is
\begin{equation}
f_{\mathbf{k}}=(-\frac{1+e^{ik_{x}}}{1+e^{ik_{y}}},0,1)^{T}.
\end{equation}%
The sublattice $B$ only vertically couples with sublattice $A$ and only
horizontally couples with sublattice $C$; the difference leads to the
momentum-dependent amplitude in $f_{\mathbf{k}}$.

In the extended Lieb lattice, the destructive interference maintains,
consequently leading to the formation of a flat band at the destructive
interference. This will be discussed in detail. The sites in sublattice $B$
are unoccupied at the destructive interference; to maintain them unoccupied,
first, additional long-range couplings $\kappa $ are applied, as indicated
by the black dotted lines in Fig. \ref{fig1}. The destructive interference
at sublattice $B$ and the flat band are unchanged; however, the form of the
eigenstates $f_{\mathbf{k}}$ in the Bloch Hamiltonian $H_{\mathbf{k}}$
alters under the additional long-range couplings $\kappa $. The
corresponding zero-energy flat band eigenstate changes to%
\begin{equation}
f_{\mathbf{k}}=(-1,0,1)^{T}.
\end{equation}%
Under the influence of the additional couplings $\kappa $, the zero-energy
eigenstate in the momentum space $(-1,0,1)^{T}$ coincides with the
eigenstate for each unit cell in the real space $\left\{
A_{j},B_{j},C_{j}\right\} =\left\{ -1,0,1\right\} $. It is worth mentioning
the significance of the formation of a destructive interference at
sublattice $B$, while sublattices $A$ and $C$ have wave functions $-1$ and$\
1$, respectively. This feature is critical for the existence of the flat
band in the non-Hermitian extension, because sublattice $A$ and $C$ can be
engineered to keep the eigenstate and destructive interference unchanged. In
this way, the flat-band energy is tunable under non-Hermitian engineering.

Maintenance of the destructive interference is the precondition for the
formation of a flat band, and the wave functions of three sublattices in the
case of flat bands ought not to be changed although some extra elements are
supplemented by adding extra gain and loss $\pm i\gamma $ respectively to
sublattices $A$ and $C$ associated with a nonreciprocal phase factor in the
coupling $Je^{\pm i\phi }$ between them. $A_{j}$ and $C_{j}$ in each unit
cell constitute a $\mathcal{PT}$-symmetric dimer \cite%
{Bender98,Ruschhaupt,NM,El,AGuo,LJin,Joglekar,CE,CHLee,Schomerus}, and $-1$
and $1$ constitute an eigenstate of the dimer at an appropriate condition
\cite{JLFB}. By acting $H_{\mathbf{k}}$ on the flat-band wave function $f_{%
\mathbf{k}}=(\psi _{A},\psi _{B},\psi _{C})^{T}$, the Schr\"{o}dinger
equations $H_{\mathbf{k}}(\psi _{A},\psi _{B},\psi _{C})^{T}=E_{\mathbf{k}%
}(\psi _{A},\psi _{B},\psi _{C})^{T}$ can be written as%
\begin{eqnarray}
E_{\mathbf{k}}\psi _{A} &=&\kappa \psi _{B}\left(
e^{-ik_{x}}+e^{-ik_{y}}+1\right) \\
&&-i\gamma \psi _{A}+Je^{-i\phi }\psi _{C},  \notag \\
E_{\mathbf{k}}\psi _{B} &=&\kappa \psi _{A}\left(
e^{ik_{x}}+e^{ik_{y}}+1\right) \\
&&+\kappa \psi _{C}\left( e^{ik_{x}}+e^{ik_{y}}+1\right) ,  \notag \\
E_{\mathbf{k}}\psi _{C} &=&\kappa \psi _{B}\left(
e^{-ik_{x}}+e^{-ik_{y}}+1\right) \\
&&+i\gamma \psi _{C}+Je^{i\phi }\psi _{A}.  \notag
\end{eqnarray}%
After substituting $(\psi _{A},\psi _{B},\psi _{C})^{T}=(-1,0,1)^{T}$ into
Schr\"{o}dinger equations, we obtain two relations $i\gamma +Je^{-i\phi
}=-E_{\mathbf{k}}$ and $i\gamma -Je^{i\phi }=E_{\mathbf{k}}$, which provide
the flat band condition
\begin{equation}
\gamma =J\sin \phi ,  \label{FB condition}
\end{equation}%
as well as the flat-band energy
\begin{equation}
E_{\mathbf{k}}=-J\cos \phi .
\end{equation}

It is evident that the adjustable flat-band energy is real. Obviously, the
real energy of the flat band is tunable by varying the coupling $J$ or the
phase $\phi $ at the appropriate match, independent of the reciprocal
inter-sublattice couplings $\kappa $. The unchanged eigenstate implies the
maintenance of the flat band. To summarize, the momentum-independent
eigenstate is essential and beneficial for the flat band in the
non-Hermitian extended Lieb lattice. To achieve the flat band in the 2D
non-Hermitian Lieb lattice, two steps are elucidated. First, through adding
extra couplings, a momentum-independent flat band eigenstate is obtained,
which is the key role that the long-range couplings $\kappa $ play in the
destructive interference. Second, the gain and loss $\pm i\gamma $ and
nonreciprocal couplings $Je^{\pm i\phi }$ are added at an appropriate match;
the eigenstate is unchanged, thereby rendering the flat-band energy tunable.

\section{Energy Bands}

\label{IV}

\begin{figure}[t]
\includegraphics[ bb=0 0 602 696, width=8.5 cm, clip]{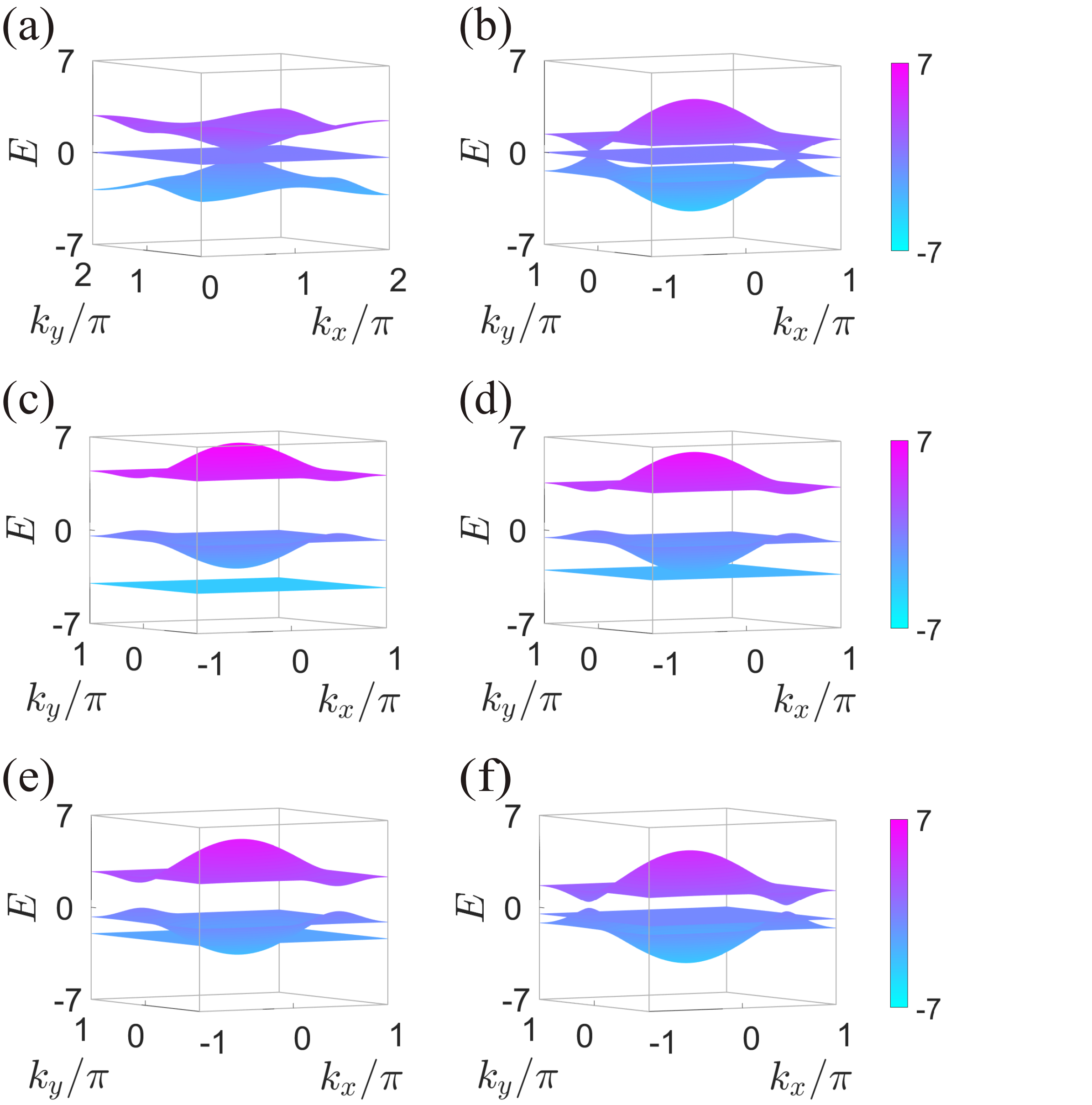}
\caption{Energy spectra of (a) the original version of the Hermitian Lieb
lattice, and (b) the Hermitian lattice with additional long-range couplings $%
\protect\kappa $. Non-Hermitian energy spectra with a flat band at (c) $J=8$
, (d) $J=6$, (e) $J=4$, (f) $J=1$ are shown; and the parameter $\protect\phi %
=\protect\pi /3$ is fixed. For all the spectra, The other coupling $\protect%
\kappa =1$ is set.}
\label{fig2}
\end{figure}

All the points of intersection of the energy spectra contain degenerate
points (DPs) and EPs. The DPs and EPs are found from the 
discriminant $\Delta =27\kappa ^{4}J^{2}s_{\mathbf{k}}^{2}\cos ^{2}\phi
-\left( 2\kappa ^{2}s_{\mathbf{k}}+J^{2}\cos ^{2}\phi \right) ^{3}$ of the
cubic equation [Eq.~(\ref{Cubic})] about the energy bands under the flat
band condition. $\gamma =J=0$ yields the Hermitian Lieb lattice. The cubic
equation of $E_{\mathbf{k}}$ is simplified to $E_{\mathbf{k}}\left( E_{%
\mathbf{k}}^{2}-2\kappa ^{2}s_{\mathbf{k}}\right) =0$. A zero-energy flat
band exists. In this situation, the discriminant $\Delta =0$ is
satisfied only when $s_{\mathbf{k}}=0$; the three bands intersect at the
zero energy; and the intersection points are DPs. Otherwise, the band
intersections are EPs in the non-Hermitian case ($\gamma \neq 0$). The discriminant $\Delta =0$\ yield the condition%
\begin{equation}
J^{2}\cos ^{2}\phi -\kappa ^{2}s_{\mathbf{k}}=0,  \label{Intersection}
\end{equation}%
provided that $s_{\mathbf{k}}\in \lbrack 0,9]$. We emphasize that $\phi
=n\pi +\pi /2$ ($n\in \mathbb{Z}$) corresponds to the case that the system
possesses the chiral symmetry. The chiral symmetry protects the existence of
a zero-energy flat band, and Eq. (\ref{FB condition}) is not necessary to be
satisfied to form the zero-energy flat band.
Although all three energies are zeros at band intersections, only two eigenstates coalesce and the two
intersections are EP2s regardless of the couplings $J$ and $\kappa $ (EP2 refers to the two-state coalescence).
For $\phi \neq n\pi +\pi /2$, the flat band
intersects a dispersive band. The EPs appear as an isolated EP or EP ring. A
single isolated EP appears at $k_{x}=k_{y}=0$ when the flat band is at the
bottom or top among the three bands; there are two isolated EPs or DPs in
the presence of chiral symmetry; and the intersections are single or double
EP rings in other cases. The EPs may disappear if $J^{2}\cos ^{2}\phi
/\kappa ^{2}$\ exceeds the value of $s_{\mathbf{k}}$. Since $s_{\mathbf{k}%
}\in \lbrack 0,9]$, $J^{2}\cos ^{2}\phi >9\kappa ^{2}$ corresponds to the
situation that EPs are absent; consequently, the three bands are separated.

We analyze the spectra of the non-Hermitian Lieb lattice. The spectrum of
the original Hermitian Lieb lattice is presented in Fig. \ref{fig2}(a).
Addition of long-range couplings $\kappa $ to the traditional Lieb lattice
does not change the energy of the flat band. The Lieb lattice is still
Hermitian and the corresponding spectrum is presented in Fig. \ref{fig2}(b).
The zero-energy points in the spectrum are two DPs. The positions of DPs in
the Brillouin zone (BZ) are fixed and independent of $\kappa $ and $\phi $.
The condition of Eq. (\ref{Intersection}) yields the existence of EPs. The
appearance of EP2s in our discussion with the nonexistence of EP3s
(three-state coalescence) indicates that the equality is satisfied. Figures %
\ref{fig2}(c)-\ref{fig2}(f) represent energy spectra for a certain constant $%
\phi \in \lbrack 0,\pi /2)$. Figure \ref{fig2}(c) is a representative
spectrum for the separable energy bands without any EP. In Fig. \ref{fig2}%
(d), the flat band and the lower dispersive band share a common point in the
BZ, which is an isolated EP2 at $k_{x}=k_{y}=0$; this requires an
appropriate match between $\kappa $ and $J$, that is, $J^{2}\cos ^{2}\phi
=9\kappa ^{2}$. If $J/\kappa $ decreases, the flat-band energy increases and
band intersection changes from a single EP to an EP ring [Fig. \ref{fig2}%
(e)] and then to two EP rings [Fig. \ref{fig2}(f)]. After that, the band
intersection ends up\ with two DPs at $J=0$. When $\phi \in (\pi /2,\pi ]$,
the flat band intersects with the upper band. The relative position of the
flat band and the dispersive bands is similar as that in the case of $\phi
\in \lbrack 0,\pi /2)$. As $J/\kappa $ increases from $0$, the flat-band
energy increases to a positive value, and the band intersections change from
two EP rings, to a single EP ring, and finally to a single isolated EP ($%
k_{x}=k_{y}=0$) at $J^{2}\cos ^{2}\phi =9\kappa ^{2}$. Subsequently,
increasing $J/\kappa $ results in the separable band structure with the flat
band being the upper band.

\section{Compact Localized States}

\label{V}
The CLSs can be found at the flat band condition Eq. (\ref{FB condition}).
The destructive interference at sublattice $B$ indicates that the wave
functions of sublattices $A$ and $C$ satisfy $\psi _{A_{m,n}}+\psi
_{C_{m,n}}=0$ for the $n$th column with the $m$th line unit cell. The Schr%
\"{o}dinger equations of the sites in the $n$th column with the $m$th line
unit cell inside the non-Hermitian Lieb lattice under the periodic boundary
condition are given by \cite{Malomed}
\begin{eqnarray}
i\dot{\psi}_{A_{m,n}} &=&\kappa \left( \psi _{B_{m,n}}+\psi
_{B_{m,n-1}}+\psi _{B_{m-1,n}}\right)  \notag \\
&&+Je^{-i\phi }\psi _{C_{m,n}}-i\gamma \psi _{A_{m,n}}, \\
i\dot{\psi}_{B_{m,n}} &=&\kappa \left( \psi _{A_{m,n}}+\psi
_{C_{m,n}}\right) +\kappa \left( \psi _{A_{m+1,n}}+\psi _{C_{m+1,n}}\right)
\notag \\
&&+\kappa \left( \psi _{A_{m,n+1}}+\psi _{C_{m,n+1}}\right) , \\
i\dot{\psi}_{C_{m,n}} &=&\kappa \left( \psi _{B_{m,n}}+\psi
_{B_{m,n-1}}+\psi _{B_{m-1,n}}\right)  \notag \\
&&+Je^{i\phi }\psi _{A_{m,n}}+i\gamma \psi _{C_{m,n}}.
\end{eqnarray}%
At the steady-state, the Schr\"{o}dinger equations reduce to%
\begin{eqnarray}
E\psi _{A_{m,n}} &=&\kappa \left( \psi _{B_{m,n}}+\psi _{B_{m,n-1}}+\psi
_{B_{m-1,n}}\right)  \notag \\
&&-i\gamma \psi _{A_{m,n}}+Je^{-i\phi }\psi _{C_{m,n}}, \\
E\psi _{B_{m,n}} &=&\kappa \left( \psi _{A_{m,n}}+\psi _{A_{m+1,n}}+\psi
_{A_{m,n+1}}\right)  \notag \\
&&+\kappa \left( \psi _{C_{m,n}}+\psi _{C_{m+1,n}}+\psi _{C_{m,n+1}}\right) ,
\\
E\psi _{C_{m,n}} &=&\kappa \left( \psi _{B_{m,n}}+\psi _{B_{m,n-1}}+\psi
_{B_{m-1,n}}\right)  \notag \\
&&+i\gamma \psi _{C_{m,n}}+Je^{i\phi }\psi _{A_{m,n}}.
\end{eqnarray}%
The solution is $\left\{ \psi _{A_{m,n}},\psi _{B_{m,n}},\psi
_{C_{m,n}}\right\} =\left\{ -1,0,1\right\} $ with the amplitudes of the
other unit cells being $\left\{ 0,0,0\right\} $. This indicates that the
CLS of the Schr\"{o}dinger equations in the real space are in accordance with the eigenstate solutions of the Schr\"{o}dinger equations in the momentum space.

The CLS has a different form in the situation where the Lieb lattice
possesses the chiral symmetry at $\phi =\pi /2$, but $\gamma \neq J$.
Under chiral symmetry, $J$ does not have to satisfy $\gamma =J\sin {(\pi /2)}$. The corresponding CLSs of the zero-energy flat band at $\gamma \neq J$
localizes in three unit cells. The representative Schr\"{o}dinger equations
for the three unit cells at the left bottom are given by%
\begin{eqnarray}
i\dot{\psi}_{A_{1}} &=&\kappa \psi _{B_{1}}-i\gamma \psi _{A_{1}}+Je^{-i\phi
}\psi _{C_{1}}, \\
i\dot{\psi}_{B_{1}} &=&\kappa \left( \psi _{A_{1}}+\psi _{C_{1}}+\psi
_{A_{2}}+\psi _{C_{2}}+\psi _{A_{3}}+\psi _{C_{3}}\right) , \\
i\dot{\psi}_{C_{1}} &=&i\gamma \psi _{C_{1}}+\kappa \psi _{B_{1}}+Je^{i\phi
}\psi _{A_{1}}.
\end{eqnarray}%
Substituting $\phi =\pi /2$ into the Schr\"{o}dinger equations, the
steady-state solutions of the non-normalized CLSs are%
\begin{eqnarray}
\psi _{A_{1}} &=&\psi _{A_{2}}=\psi _{A_{3}}=-1,  \label{Psi1} \\
\psi _{B_{1}} &=&i\left( J-\gamma \right) /\kappa ,\psi _{B_{2}}=\psi
_{B_{3}}=0,  \label{Psi2} \\
\psi _{C_{1}} &=&\psi _{C_{2}}=\psi _{C_{3}}=1.  \label{Psi3}
\end{eqnarray}%
Notably, similar structure of CLSs appear at other corners and inside the
non-Hermitian Lieb lattice.

\begin{figure}[t]
\includegraphics[ bb=0 0 596 213, width=8.7 cm, clip]{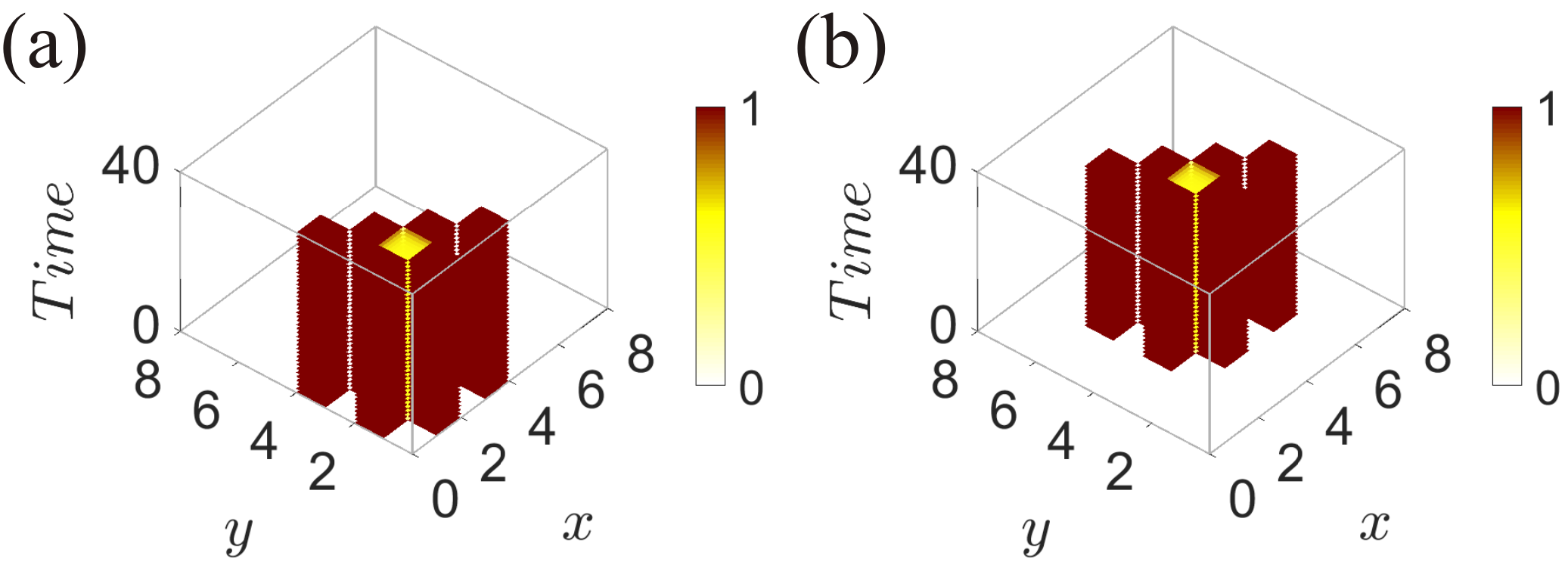}
\caption{Time evolution $\left\vert \langle \Phi \left( t\right) \left\vert
\Phi \left( t\right) \right\rangle \right\vert ^{2}$ of CLSs at (a) the
lattice boundary and (b) inside the lattice. The
unit of time is $1/\protect\kappa $. The initial excitations are in
the form of Eqs.~(\protect\ref{Psi1}) to~(\protect\ref{Psi3}). The Lieb
lattice is chiral symmetric under system parameters $J=1$, $\protect\kappa %
=1 $, $\protect\phi =\protect\pi /2$, $\protect\gamma =1/4$.}
\label{fig3}
\end{figure}

The time evolution of CLSs is obtained
\begin{equation}
\left\vert \Phi \left( t\right) \right\rangle =e^{-iHt}\left\vert \Phi
\left( 0\right) \right\rangle .
\end{equation}
The initial excitations $\left\vert \Phi \left( 0\right) \right\rangle $ are
chosen as when the CLSs are at the lattice boundary and inside the lattice.
The numerical simulations of the dynamics are shown in Fig. \ref{fig3},
where the excitation intensity $\left\vert \langle \Phi \left( t\right)
\left\vert \Phi \left( t\right) \right\rangle \right\vert ^{2}$ is depicted.
In the simulation, we choose eight sites both in $x$ and $y$ directions. The
initial excitation located at the edge of the lattice in Fig.~\ref{fig3}(a)
is provided by Eqs.~(\ref{Psi1}) to (\ref{Psi3}) with $\left\vert \Phi
\left( 0\right) \right\rangle =(-1,3i/4,1,-1,0,1,-1,0,1)^{T} $. In Fig. \ref%
{fig3}(b), the initial excitation locates inside the lattice. The
non-excited zero intensities in both two numerical simulation diagrams are
displayed in white. We notice that the profiles of the initial excitations
are unchanged in the whole dynamical process, being confined and
diffractionless. The time evolution dynamics verify that the CLSs are
localized in three unit cells in the non-Hermitian Lieb lattice under chiral
symmetry $\phi =n\pi +\pi /2$ ($n\in\mathbb{Z}$), but $\gamma\neq J\sin\phi$.

\section{Tasaki's decorated square lattice and the dice lattice}

\label{VI} A flat band can be realized in other 2D non-Hermitian lattices by
employing the approach introduced in Sec. \ref{III}. In this section, as
examples, we apply the two steps to propose a tunable flat band in the
Tasaki's decorated square lattice, dice lattice and the kagome lattice. In
the first step, we propose a 2D lattice possessing a momentum-independent
flat band for the lattice in the momentum space; in the second step, we
apply additional terms involving non-Hermitian elements to manipulate the
flat-band energy without changing the eigenstate of the flat band.

\begin{figure*}[thb]
\includegraphics[ bb=-20 0 604 378,width=17.5 cm, clip]{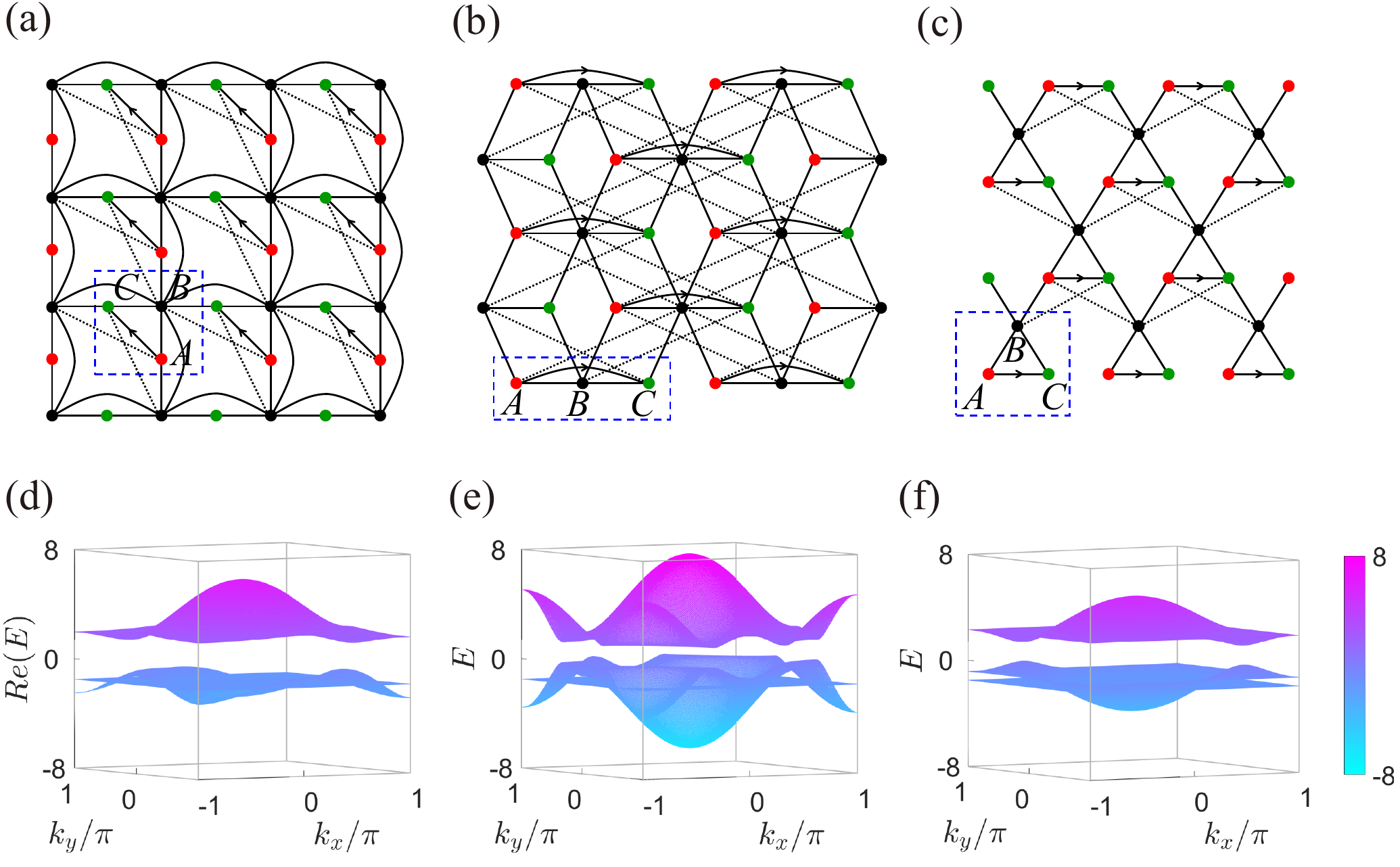}
\caption{Schematics of the 2D non-Hermitian (a) Tasaki's decorated square
lattice, (b) dice lattice, and (c) kagome lattice; and their corresponding
real part of the energy spectra (d), (e) and (f). The unit cells of all the
lattices, indicated in the blue dashed boxes, contain three sites. The
legends are identical to those shown in Fig.~\protect\ref{fig1}. All the
band energies in the spectra are real numbers except for the two dispersive
bands in (d). In the plots (d)-(f), the parameters are $\protect\phi =
\protect\pi /3$, $\protect\kappa =1$, and $J=3$.}
\label{fig4}
\end{figure*}
The Tasaki's decorated square lattice is schematically illustrated in Fig. %
\ref{fig4}(a), which is a 2D Lieb lattice with extra long-range couplings of
four nearest-neighbor sites for sublattice $B$. Applying the same approaches
elucidated for the aforementioned Lieb lattice, the momentum-independent
eigenstate is achieved for the flat band similar to that for the Lieb
lattice. In addition, the gain and loss associated with the nonreciprocal
couplings added in the Tasaki's decorated square lattice similar to the 2D
non-Hermitian Lieb lattice maintain the flat band, and the flat-band energy
and eigenstate are identical to those of the non-Hermitian Lieb lattice
previously discussed in Sec. \ref{II}. The tunable flat band energy $-J\cos
\phi $ is depicted in Fig. \ref{fig4}(d).

Furthermore, the non-Hermitian dice lattice depicted in Fig. \ref{fig4}(b)
supports a tunable flat band. The configuration of coupling between
sublattices $B$ and $C$ is not symmetric with respect to that between
sublattices $B$ and $A$, which results in the momentum-dependent eigenstate
for the zero-energy flat band of the dice lattice. This indicates that extra
couplings are required to be added in the original dice lattice to obtain
the momentum-independent eigenstate of the flat band. The additional
couplings are represented by the black dotted lines. Then, the couplings
between sublattices $B$ and $A$ are equal to those between sublattices $B$
and $C$. In the second step, with the appropriate match between the added
nonreciprocal coupling and the non-Hermitian gain and loss, the flat band is
formed at the destructive interference. The corresponding flat band energy
in Fig. \ref{fig4}(e) is identical to that in Fig. \ref{fig4}(d).

In Fig. \ref{fig4}(c), the method to achieve tunable flat bands in
non-Hermitian systems is applied for the kagome lattice. For the original
kagome model, the flat band is the lowest energy band with all couplings
being positive; the flat-band energy is nonzero, which differs from that in
the Lieb, Tasaki's decorated square, and dice lattices. The couplings
between sublattices $A$ and $C$ are no longer zero in this case, thus making
the first step to form the momentum-independent eigenstate of the flat band
different; eliminating the couplings $\kappa $ between sublattices $A$ and $%
C $ solves the problem. The second step
in achieving the flat band is similar to that for the aforementioned 2D
lattices. The spectrum is depicted in
	Fig. \ref{fig4}(f), the flat band energy is identical to those in Figs. %
	\ref{fig4}(d) and \ref{fig4}(e).

\section{Summary}

\label{VII}

{In this study, we propose a method to realize a flat band in a 2D
non-Hermitian Lieb lattice, whose energy is flexible by the interplay of the
non-Hermiticity and synthetic magnetic flux. We propose two steps to achieve
a 2D non-Hermitian lattice that supports a flat band. In the first step, we
obtain the momentum-independent eigenstate of the flat band for the Bloch
Hamiltonian, which corresponds to CLSs in its unit cell. In the second step,
a nonreciprocal coupling and non-Hermitian gain and loss are introduced at
appropriate matches to maintain the destructive interference. The flat-band
energy is tunable for the different matches between the coupling and the
non-Hermitian gain and loss. }Furthermore, we apply our approach to other 2D
lattices. The proposed non-Hermitian Tasaki's decorated square lattice and
the dice lattice are demonstrated to support tunable flat bands. We provide
a new perspective in the generation of the flat bands, and our findings are
applicable to a vast range of 2D Hermitian lattices and beyond for achieving
flat bands in non-Hermitian lattices.

\section{ACKNOWLEDGMENTS}

This work was supported by National Natural Science Foundation of China
(Grants No.~11975128 and No.~11605094), and the Fundamental Research Funds
for the Central Universities, Nankai University (Grants No.~63191522 and
No.~63191738).

\end{document}